# Efficient Two-Level Scheduling for Concurrent Graph Processing

Jin Zhao

**Abstract**—With the rapidly growing demand of graph processing in the real scene, they have to efficiently handle massive concurrent jobs. Although existing work enable to efficiently handle single graph processing job, there are plenty of memory access redundancy caused by ignoring the characteristic of data access correlations. Motivated such an observation, we proposed two-level scheduling strategy in this paper, which enables to enhance the efficiency of data access and to accelerate the convergence speed of concurrent jobs. Firstly, correlations-aware job scheduling allows concurrent jobs to process the same graph data in Cache, which fundamentally alleviates the challenge of CPU repeatedly accessing the same graph data in memory. Secondly, multiple priority-based data scheduling provides the support of prioritized iteration for concurrent jobs, which is based on the global priority generated by individual priority of each job. Simultaneously, we adopt block priority instead of fine-grained priority to schedule graph data to decrease the computation cost. In particular, two-level scheduling significantly advance over the state-of-the-art because it works in the interlayer between data and systems.

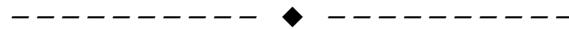

## 1 INTRODUCTION

Mny enterprises, such as Twitter, Facebook, etc., widely use graph processing to analyze data due to graph enables clearly express the inherent interdependencies between entities (i.e., social networks and transportation graphs). Therefore, in recent years, related research has attracted extensive attention from academia and industry. For processing graph efficiently, there are a great deal of graph computing platforms were proposed. The distributed systems exploite the powerful computation resources of clusters to process large graphs, such as Pregel [6], GraphLab [10], PowerGraph [8], GraphX [9], etc. Alternatively, single-machine systems, like GraphChi [11], X-Stream, Gridgraph, etc., are able to handle graphs with billions of edges by using secondary storage efficiently, and largely eliminate the challenges of synchronization overhead [5, 7] and load imbalance [4] in distributed frameworks.

However, with the rapidly growing demand of graph processing in the real scene, the number of concurrent graph processing jobs is greatly increased in data analysis platforms. Such as, Didi data analysis platform need to do more than 9 billion Route Planning daily in 2017, in other words, it is about 6 million times per minute. Therefore, it's important that handle concurrent jobs more efficiently. Based on the feature that most concurrent jobs occur on the same graph structure data, the main work of concurrent jobs includes Seraph [1] which tries to solve the problems of inefficient use of memory and high cost of fault tolerance through decoupling the data model and new computing logics. So, it means that multiple jobs are able to share the same graph structure data in memory, by it improve the utilization of computer resources. There is an important characteristic is **data access correlation** which caused by jobs share same graph structure data in memory. Such as, different concurrent jobs access the same nodes (or edges) in graph data. However, existing graph processing systems does not utilize this feature, resulting in accessing data in memory inefficiently.

Firstly, as we known the *poor locality* which is attributed to the random accesses in traversing the neighborhood nodes in graph processing. What's worse is that the *locality* will be even worse in concurrent jobs, which is because of jobs access data independently and further causes CPU to repeatedly request the same data in memory at different moment for different jobs. In addition, the times of repetitively access graph data goes up as the number of jobs increases. For two different concurrent jobs, they will lead CPU to access the same nodes (or edges) twice in one iteration altogether. Hence, in the conventional hierarchical memory system, this challenge will lead to significant multitudes of data movements. It further gives rise to low performance of data accessing and inefficiency of Cache using. We called this challenge is **Memory Access Redundancy**.

Secondly, iterative computation is widely used in graph processing. Although some work, such as PrIter [2], can accelerate the convergence speed of iterative algorithms by prioritized execution, they incur Memory Access Redundancy and more random accesses when processing multiple concurrent jobs. For example, multiple jobs have some intersections in priority queues and access them separately, it will bring about Memory Access Redundancy. Moreover, the computation cost of the fine-grained priority (each node has a priority) can be decreased by replaced with slightly coarse-grained priority.

In this paper, according to the characteristic of data access correlation, we try to slove the above problems by two kinds of effective scheduling policies: **correlation aware job scheduling (CAJS** for short**)** and **multiple priority data scheduling (MPDS** for short**)**. First of all,

CAJS transfers a part of graph data into Cache, and dispatches jobs to process this graph data through the convergence situation. Hence, in one iteration, CPU just need request the data in memory once, then accesses this data in Cache. So, it can greatly alleviate the challenge of CPU repeatedly requests the same data in memory and improve the efficiency of the data access in conventional hierarchical memory systems. Secondly, in order to accelerate the convergence speed of concurrent jobs, MPDS decides which parts of graph data need to transfer into Cache through global priority generated by individual priority of each job. Simultaneously, MPDS selects some parts of graph data to process via block priority instead of fine-grained priority, and we further propose a dual-facters order (DO) algorithm to extract the block priority queues of each job fastly.

In summary, there are two scheduling strategies which are used to optimize the data access in memory and accelerate the convergence speed of jobs, thus improving the throughput of the systems. Therefore, this work's main contributions are as follows:

- We present a novel job scheduling scheme (CAJS), which can reduce the times of CPU directly access memory by proactively dispatching concurrent jobs to process graph data in Cache.
- We design a slightly coarse-grained priority strategy (MPDS) to accelerate the convergence speed of concurrent jobs. To make the computation of priority inexpensive, we adopt block priority in prioritized execution. The DO algorithm, proposed in MPDS, enable fastly extract the blocks which have higher individual priority in most jobs, then dispatching concurrent jobs to process those blocks by CAJS.

The rest of this paper is organized as follows. Section 2 introduces the practical issue of memory access redundancy and slow convergence in concurrent jobs. In Section 3, we introduce the main idea and framework of this work. Sections 4 respectively introduce the two strategies in detail. Then we describe related work in Section 6. Finally, we conclude this paper in Section 7.

## 2 MOTIVATION

The current graph processing systems are basically designed and implemented for a single job. But in practice, with the rapid development and wide application of Internet technology [3], the demand of data analysis is greatly increased. Therefore, the number of jobs to be run on the same platform is greatly increased, it easily leads to jobs overlapping in time (i.e., concurrency). To illustrate this, we collect one-month (Nov 1, 2017) workload from a social network company. Figure 1 depicts one week's workload (number of jobs at each time) of graph computation. The stable distribution shows that a significant number of jobs are executed concurrently every day. At peak time, there are more than 20 jobs submitted to the platform. We also depict the complementary cumulative distribution of the number of concurrent jobs in each time period (one second) in Figure 2. As it shows, more than 83.4% of time has at least two jobs executed concurrently. The average number of concurrent jobs is 8.7.

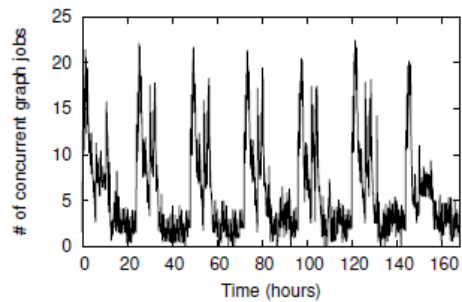

Fig. 1. One week's workload of graph computation in real social nework

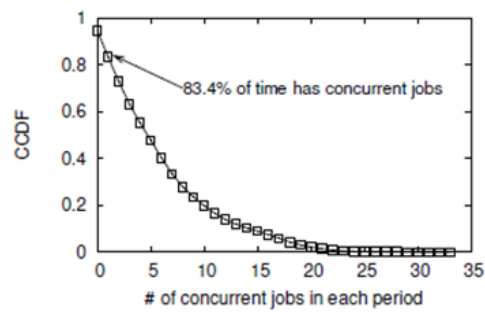

Fig. 2: The distribution of graph jobs running on one social network

Recently, the main work of concurrent jobs includes Seraph [1] which proposed a decoupling the data model to support multiple jobs sharing the same graph structure data in memory and adopts a copy-on-write semantic to isolate the graph mutation of concurrent jobs.

Although existing work can solve the problem of inefficient use of memory, some challenges result in accessing memory inefficiently. Simultaneously, if the convergence speed of concurrent jobs can be accelerated effectively, the computational performance and throughput of the graph processing systems will be greatly improved.

### 2.1 Memory Access Redundancy

Although multiple jobs sharing the same graph structure data in memory, due to all jobs need to traverse most of the edges, nodes, as well as weights in memory and each job running independently, the same data in memory will be accessed by CPU at different times with different jobs. Hence, it will cause the memory access redundancy, and unnecessary memory bandwidth occupation. At the same time, the CPU cache performance will be decreased, and reduced the speed of accessing data.

The current mode of concurrent jobs access graph structure data in memory is shown in Figure 3. And it represents that jobs access data in memroy at different times (T1, T2, and T3) in one iteration. At T1, when Job1 accesses D2, the rest of jobs (e.g., Job2, Jobn, etc,.) may be access other data blocks. Therefore, the D2 will be transferred to the Cache line of ID as j to facilitating CPU access D2. After a while (T2), Jobn accesses Di, and Di is mapped to Cache line of ID as j. So D2 in the Cache line will be replaced by Di. But Job2 may be acess D2 at T3. However, CPU can not access D2 in Cache because of D2 has been replaced by Di. So D2 needs to be reloaded into the Cache. We can clearly realize that D2 was copied from main memory to cache tiwce in this mode. Hence, the CPU Cache hit ratio has been reduced.

Apparently, with the number of concurrent jobs increases, the number of same data in memory were accessed by CPU at different times will be increased in practice. Simultaneously, we record the change of Cache miss rate (as shown in Figure 4) and the percentages of CPU cache stall (the CPU time waiting for the requested data due to the cache miss) and the CPU execution time while the CPU keeps computing (as shown in Figure 5) as the number of concurrent jobs increased. We can realize that tht performance of Cache was decreasing because of Cache hit rate and the times of Cache miss were increasing. A very important reason is that the same data is accessed by different jobs at different time, resulting in the same data transfered to Cache many times. Therefore, in this work, we want to propose a method to reduce the negative impacts by reduce the number of times of copying data from main memory to cache, or from slower cache to faster cache. We will introduce this method clearly in Section 4.

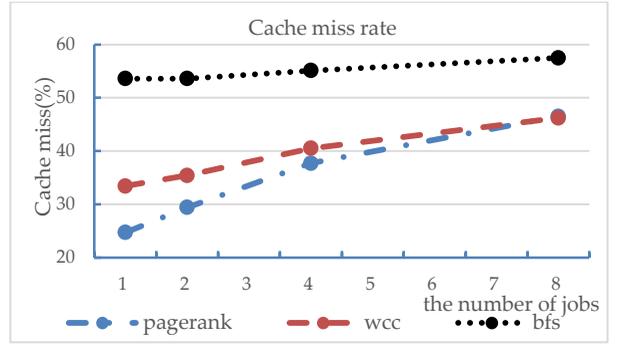

Fig. 4: Cache miss rate as the number of concurrent jobs increased.

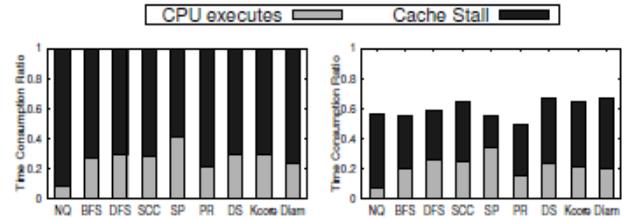

Fig. 5: CPU execution and CPU Cache Stall over sd1-arc

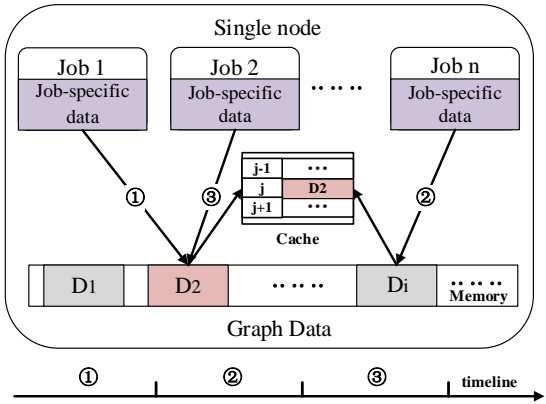

Fig. 3: Current mode of data access. (①,②,③ are represent T1, T2, T3.)

## 2.2 Slow Convergence

Iterative computation is a universal method in big data mining algorithms, such as social network analysis, recommendation systems, and so on. Hence, the existing graph processing algorithms basically adopt iterative computations. Therefore, it is essential to speed up the convergence speed in iterative computations of large-scale data mining algorithms. PrIter [2] enables fast iterative computation by providing the support of prioritized iteration, it means extracts the subset of data that have higher priority values to perform the iterative updates. Importantly, priority value of each node is determined by the influence on convergence (e.g., $\Delta P_j^k$ in PageRank).

However, the existing methods only suitable for accelerating the convergence speed of a single job, but not for concurrent jobs. Because the algorithm characteristics and computation states are different of concurrent jobs, the priority values of jobs are not identical. Such as, there are three concurrent jobs: Job1, Job2, and Job3. It is assumed that priority queue for each job are {A, B, C, D}, {B, C, D, E}, {C, D, E, F}, respectively. Hence, CPU need to access nodes {A, B, C, D, E, F} for concurrent jobs, and node C may be transfered into Cache three times. Hence, based on the priority values of each job to prioritize iteration separately will cause most of the data in memory be accessed and make the prioritized iteration inefficient.

In particular, it's usually used to save memory resources by sharing graph data in memory for concurrent jobs. In single-machine graph processing systems, if a job has completed the prioritized iteration in one interation firstly, the finished job need to wait for other jobs to finsh the prioritized iteration before the new graph data can be transferred into memory. So, if the finished job continues to compute other nodes which have low priority values when waiting, we believe the job will converge more quickly. At the same time, because it is necessary to load graph data partitions to memory frequently in single machine systems and the number of iterations will increase in prioritized iteration, it is another very important problem that the secondary storage I/O is slow.

So, we want to search the priority values which are suitable for whole concurrent jobs and solve the problem of the secondary storage I/O is slow in single machine to accelerate the convergence speed with the least amount of memory access.

## 3 APPROCHES OVERVIEW

To fully utilize the characteristic of data access correlation, we present a novel concurrent jobs schedule model to improve the efficiency of data access in memory. And based on this model, we farther propose a slightly coarse-grained strategy of prioritized iteration to accelerate the convergence speed of jobs. These two strategies can work together to improve the throughput of the graph processing systems for concurrent jobs.

In nature, the low performance of memory access is caused by frequent transfer data from memory to Cache, or from slower cache to faster cache. As shown in Fig. 4, the traditional memory access model of concurrent jobs will make memory access inefficient.

Hence, to efficiently slove this problem, we present a neoteric slightly coarse-grained memory access model of concurrent jobs. The mode of execution is jobs access the same data in Cache simultaneously to reduce the number of the same data write into the Cache.

In detail, suppose there are N concurrent jobs, we named it $J_1$ to $J_N$. In the process of computation, the concurrent jobs will access graph structure data for iteration computation. Hence, if part of the graph data has placed in Cache, we can schedule jobs to process the graph data in Cache according to convergence condition of every jobs to improve data access efficiency. In another word, we first schedule some graph data to the Cache, and then dispatch the job (e.g. $J_i$, i ∈ [1, N]) that are not converging on the data to calculate the data.

Furthermore, prioritized iteration enables accelerate iterative computation by processing the subset of data with higher priority values, so which part of the graph structure data need to be sequentially transferred into Cache for computation has a great impact on convergence speed of jobs. Therefore, we generate a global priority values based on the different data priority values in each job, and transfer graph data to Cache according to the global priority values to accelerate iterative computation of multiple jobs.

Note that we schedule the data in **blocks**, and a block can be placed in the Cache. So, we can suppose that the graph data is consisted of X blocks, that are, $B_1$ to $B_X$. We use blocks to schedule data for the purpose of reduce the cost of the priority values maintenance. Because of the graph structure data may be very large (e.g. a graph with billions of vertices), if fine-grained scheduling is carried out in nodes, it will cause huge maintenance costs of priority values and convergence condition of every jobs.

The Concurrent Processing Strategies can be divided into two steps: *1)* transfer a block, $B_i$, from memory to cache by a priority data schedule strategy; *2)* schedule job to process the block of $B_i$ simultaneously according to convergence condition of every jobs. These two steps are two functions attached to the traditional graph processing systems, which makes the systems more efficient for concurrent jobs.

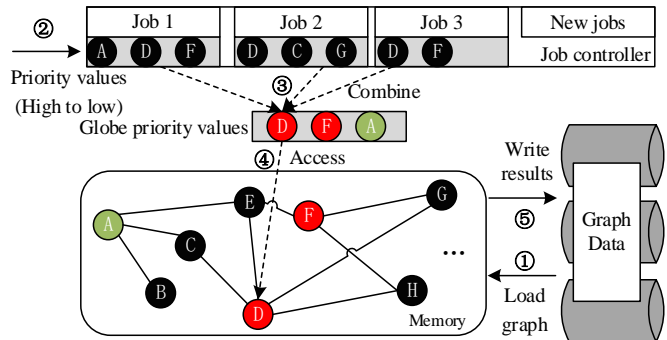

Fig. 6: Workflow of Methods Framework

## 4 TWO-LEVEL SCHEDULING STRATEGIES

In this section, we will describe in detail the design of the two strategies and the collaboration between them. Specifically, all of the strategies proposed in this paper are based on sharing graph data in memory for concurrent jobs.

### 4.1 Approches Framework

To improve throughput of the graph processing systems in concurrent jobs, we propose two strategies according to fully utilize the characteristic of data access correlation: **convergence aware job scheduling** and **multiple priority data scheduling**. These two strategies enable efficient enhance the efficiency of memory access and accelerate the convergence speed of concurrent jobs, and also are an improvement on the traditional graph processing systems. Fig. 6 shows workflow of approches framework in a single node. It can also be applied to distributed systems by using these two strategies to multiple nodes in distributed environments. The following is described in detail.

There is no doubt that graph processing systems first need to load the graph data into memory (step ①). Secondly, systems need to calculate the priority values of graph data for each job, and the length of priority queue of each job is the same. In particular, the priority values of graph data are set to the same in the first iteration. (step ②). At step ③, systems combine the global priority values that satisfy all concurrent jobs through the priority values in the priority queue of each job. Then, systems transfer graph data form memory to Cache by global priority values, and schedule jobs to access data in Cache for computation through the Job Contruller (step ④). Finally, the job results will be witten to secondary storage during the job execution (Step ⑤). When a new job is dispatched to Job Controller, a new priority values are created to join the Concurrent Processing Strategies. According to the description, multiple priority data scheduling is corresponding to step ② and step ③ (introduced in Section 4.2), and convergence aware job scheduling is matching to step ④ (detailed in Section 4.3).

## 4.2 Multiple Priority Data Scheduling

To efficient accelerate the convergence speed of concurrent jobs, we adopt a similar method with PrIter, but we have improved this method. At first, we take block-based approach to slightly coarse-grained scheduling graph data to reduce the overhead of the priority values state maintenance. Then we use global priority values to schedule graph data in prioritized iteration, and avoid the problem of low overall performance by each job scheduling data according to their own priority values.

In order to efficiently achieve these improvements, we have to solve some problems caused by these improvements. For example, how to accurately and effectively determine the priority values of the block, etc. So, next we will show how to solve these problems in detail.

### 4.2.1 Individual Priority Blocks

Since the global priority values are generated based on the Individual Priority Values, we need to first determine the priority values of each job and calculate priority values in blocks.

Obviously, for a job, if the more nodes are not converged in a block, the greater the effect of the block on the convergence speed of the job, and it should be given a higher priority. Secondly, different nodes have different influence on the convergence speed of the jobs. For Pagerank, in an iteration, the larger the Pagerank value changes, the greater the effect of the convergence speed, and the greater the priority value of the node. But for SSSP, Node j is eligible for the next iteration only if D(j), the distance to node j, has changed since the last iteration on j. Priority is given to the node j with smaller value of D(j). Hence, it is necessary to determine the priority value of the block according to the number of unconvergent nodes in the block and the algorithm characteristics.

Firstly, the priority value needs to be represented by an exact number. So, the priority value of each node is decided by a function *De_In_Priority*, which is for users to specify each node's enforcement priority in consideration of the algorithm characteristics. For example, in SSSP, the priority value is the negative value of the shortest distance of each node, while in PageRank, the priority value is exactly the same value as the $\Delta_R$. Secondly, after calculating the priority value of nodes in a block, *De_In_Priority* further calculates and compares the priority values of nodes to obtain the order of blocks' priority values, the following will describe how it works.

The priority value of block is determined by two aspects: the number of unconvergent nodes and the priority value of each unconvergent node. So, the priority value of block can be expressed as the $<\text{Node}_{un}, \overline{P_{value}}>$ pair. Detailedly, $\text{Node}_{un}$ is the number of unconvergent nodes, and $\overline{P_{value}}$ is the average of the node's priority value is not converged in the block. The formula can be described as,

$$\overline{P_{value}} = \sum_{i=0}^{N} P_i / \text{Node}_{un} \quad (1)$$

where N is the number of node in block, $P_i$ is the priority value of node i, if the node i has been conversed, $P_i = 0$.

TABLE 1
COMPARE TWO BLOCKS' PRIORITY

|  | Possible Scenarios |  | Result |
|---|---|---|---|
| case 1 | $\overline{P_{value_a}} > \overline{P_{value_b}}$ | $\text{Node}_{un_a} > \text{Node}_{un_b}$ | $P_a > P_b$ |
| case 2 | $\overline{P_{value_a}} > \overline{P_{value_b}}$ | $\text{Node}_{un_a} < \text{Node}_{un_b}$ | ? |
| case 3 | $\overline{P_{value_a}} = \overline{P_{value_b}}$ | $\text{Node}_{un_a} > \text{Node}_{un_b}$ | $P_a > P_b$ |
| case 4 | $\overline{P_{value_a}} > \overline{P_{value_b}}$ | $\text{Node}_{un_a} = \text{Node}_{un_b}$ | $P_a > P_b$ |

$\overline{P_{value_a}}$ *represents the average priority value of $B_a$, and* $\text{Node}_{un_b}$ *is the number of nodes that are not convergent in $B_b$, $P_a$ represents the priority value of $B_a$.*

Why we express pair like this? We can consider that the more nodes without convergence, and the higher the priority value of each node, the higher the priority value of the block will be. Therefore, the priority values of blocks need to weigh these two aspects effectively. Such as, if we use the total priority value to express the priority value of the block, because it can reflect the number of unconvergent nodes in the block. But it is not effectively because if $B_a$ only has a small number of unconvergent nodes with high priority values and $B_b$ has lots of unconvergent nodes with low priority values which will lead to total priority value of $B_a$ less than $B_b$. However, $B_a$ should be given a higher priority because of $B_a$ has a small amount of computation. According to this problem, we can solve it by comparing the average of the node's priority value, because $B_b$ must has a low average value. But there is another problem that is if $B_c$ and $B_d$ have average values of little difference ($\overline{P_{value_c}} < \overline{P_{value_d}}$), but $B_c$ has more unconvergent nodes than $B_d$. Obviously, $B_c$ should be given a higher priority by comparing the total priority value. Hence, the priority value of the block can be reflected by means of $\overline{P_{value}}$ and total priority value.

Through the above discussion, we propose a dual-facters order (DO) algorithm in *De_In_Priority* to accurately determine the priority value of the block. The rationale of the DO algorithm is to first compare the priority of two blocks based on $\overline{P_{value}}$. Secondly, if the $\overline{P_{value}}$ of the two blocks vary within a certain range (i.e., $|\overline{P_{value_c}} - \overline{P_{value_d}}| < \varepsilon$, $\varepsilon$ is a constant or something), the the priority of two blocks is based on the total priority value ( $\text{Node}_{un} \times \overline{P_{value}}$ ). And during the calculation of blocks, instead of a pass over the entire blocks as an iteration, a pass through a selected subset as a subpass is performed based on the entries' priority values. A subset is called a priority queue, and the size of the priority queue is assumed to be q. The size of the q will have a great impact on the performance of this method, which will be discussed in detail later. Hence, the DO algorithm also need to select the subset of blocks with higher priority values in each iteration simultaneously. More details on this algorithm will be discussed later in this section.

### 4.2.2 The DO Algorithm

We give an algorithm to obtain an ordered priority queue for each job to effectively generate global priority queue. How to compare the priority value of blocks is the key to generate orderly priority queues of jobs, and we adopt the <$Node_{un}$, $\overline{P_{value}}$> pair to slove it. For the comparison of two block priorities, there are four possible scenarios as shown in Table 1, and suppose that two blocks are $B_a$ and $B_b$. In case 1, 3, 4, the priority value of $B_a$ is obviously higher than $B_b$. But in case 2, we need to discuss when the difference between the priority values of the two blocks is small, and this problem and solution has been described above. In this work, we set ε = 0.2 × $\overline{P_{value_a}}$. The priority comparison method of the two blocks is shown in Function 1. Therefore, we can add Function 1 to the sorting algorithm to prioritize the blocks of each job.

**Fuction 1**: Compare two blocks' Priority(CBP)
  **Input**: $Pair_a$ = < $Node_{un_a}$, $\overline{P_{value_a}}$ >, $Pair_b$ = <$Node_{un_b}$, $\overline{P_{value_b}}$>
  **Output:** Is the priority of $B_a$ higher than $B_b$?
1   state ← True;
2   if $\overline{P_{value_a}}$ < $\overline{P_{value_b}}$ then
3     swap ($Pair_a$, $Pair_b$);
4     state ← ¬ state;
5   end
6   if $Node_{un_a}$ < $Node_{un_b}$ then
7     if $\overline{P_{value_a}}$ − $\overline{P_{value_b}}$ < 0.2 × $\overline{P_{value_a}}$ &&
8       $\overline{P_{value_a}}$ × $Node_{un_a}$ < $\overline{P_{value_b}}$ × $Node_{un_b}$ then
9         state ← ¬ state;
10    end
11  end
12  return state

The DO algorithm should select the top q blocks with the highest priority in each iteration. But we don't want to sort the whole blocks because it's time consuming. Furthermore, it is unnecessary to select the exact q top priority blocks. Hence, wo adopt a method of approximate selection as shown in Fuction 2. The idea of this heuristic is that, the distribution of the priority values in a small (default 500) samples set reflects the distribution of priority values in whole blocks. By sorting the samples in the descending order of the priority values, the lower bound of the priority value of the top q records can be estimated to be the $(q * s/B_N)$th record's priority value in the sorted samples set. Then we only need to prioritize the extracted q blocks.

Through the above introduction, we can calculate the time complexity of the DO algorithm, and can can be expressed as: O ($B_N$) + O (q log q), among them, q = C ∗ $B_N/\sqrt{V_N}$ = C ∗ $\sqrt{V_N}$ / $V_B$, $V_B$ is the number of nodes in a block (discussing in section 5.1). The further time complexity is:
  O ($B_N$) + O (($\sqrt{V_N}/V_B$) log( $\sqrt{V_N}$ /$V_B$)) <
      O ($B_N$) + O ($V_N/V_B$) = O ($B_N$) + O ($B_N$) .    (2)
Through this approximation, we just only take O ($B_N$) time on extracting the top priority blocks instead of O ($B_N$ log $B_N$) time.

### 4.2.3 Global Priority Blocks
The individual priority queue of blocks can be obtained by the fanction of De_In_Priority. In order to achieve uniform prioritized iteration, we need to generate the global priority queue based on the priority queue of each job.

Different from individual priority blocks only consider the impact of a block's priority value on the convergence speed of an individual job, global priority blocks also consider the interplay of the priority value of blocks of each job, such as, the number of times a block occurrence

**Fuction 2**: The DO algorithm
  **Input**: Pair Table of each job Ptable, number of blocks $B_N$, length of queue q, samples set size s
  **Output:** priority queue queue
1   samples ← randomly select s pairs from Ptable;
2   sort samples in priority-descending order;
3   cutindex ← $q * s/B_N$;
4   thresh ← samples [cutindex];
5   i ← 0;
6   foreach pair r in Ptable do
7     if CBP (r, thresh) then
8       queue [i] ← r;
9       i ← i + 1;
10    end
11  end
12  sort queue in priority-descending order;

in each job priority queue. To make the prioritized iteration more efficient, we also set the global priority queue length q.

The global priority queue is generated by a function *De_Gl_Priority*, which shows the trade-off between the gain from the global convergence speed and the cost from the individual convergence speed. This is because of a block that can effectively accelerate the convergence speed of most jobs may not be the most effective way to accelerate the convergence speed of some jobs. Therefore, in addition to considering the block that has an impact on the convergence of the overall jobs, it is also necessary to consider the block that has a greater impact on the convergence of an individual job. This may happen that the highest priority block in the priority queue of an individual job may be has little impact on accelerate the convergence speed of most other jobs, causing it's not in the globe priority queue. Therefore, we need to add these blocks to the global priority queue.

The process of synthesizing global priority queue is illustrated by an example, as shown in Fig. 7. The Pri of each priority queue is assigned from q to 1, so we accumulate the Pri of each block in whole priority queues as the priority value for the global priority queue, e.g., for $B_d$ is 2q-1. Therefore, we can use a method similar to the Fuction 2 to approximatively extract the block with high priority value. Note that some of the space of global priority queue is reserved for some blocks which have higher priority values in individual job, and the cumulative sum of the Pri of those blocks isn't higher. Hence, we set a threshold α (α ∈ (0,1]) to determine the allocation of space, and set the α default to 0.8.

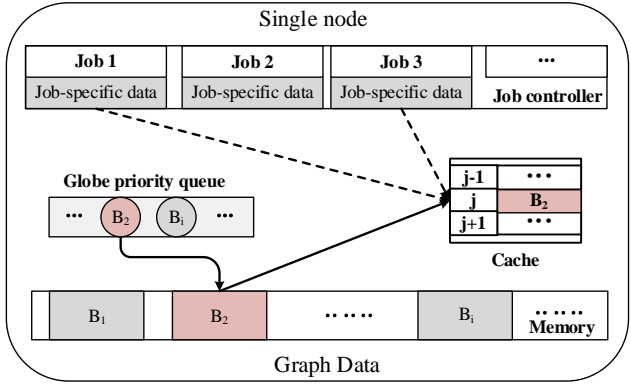

Fig. 8: Concurrent access model

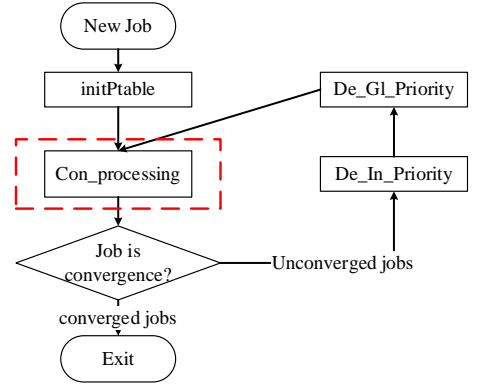

Fig. 9: The use of APIs.

Once the priority queue of every jobs is given, *De_Gl_Priority* first approximatively selects the α ∗ q top priority blocks with the high cumulative sum of the Pri to add to global priority queue, and then add the blocks which have higher priority values in individual job, but which blocks not in the globe priority queue. Next, the systems need to schedule jobs processing graph data based on globe priority queue.

### 4.3 Convergence Aware Job Scheduling

To improve the efficiency of jobs access graph data in memory, we propose a mode of concurrent access by convergence aware. The detail is that first to put the blocks with higher globe priority value into Cache, and then the job controller will dispatch the corresponding jobs to process the block, which can effectively reduce the number of the Cache pollution.

The concurrent access model is shown in Fig. 8. Such as $B_2$ has a higher priority, so we transfer $B_2$ from memory to Cache, then the job controller will dispatch the corresponding jobs (e.g. Job1, Job3, etc.) to process $B_2$. Therefore, the efficiency of concurrent jobs access graph data will be improved.

We adopt a simple and efficient method to transfer blocks from memory to Cache, that is, using the Cache mechanism of the operating system. When graph processing system need to transfer a block to Cache, it can dispatch a job to access this block. If this block is not in Cache, it will be transfer to Cache by the operating system. Then, the job controller schedule jobs which unconverged in this block to processing. Therefore, the job controller need to know the convergence of each job in each block. In particular, the convergence can be gained through priority queue of each job.

### 4.4 Fuction API

The approachs presented in this work has several programming interfaces supplied to users for accelerating graph processing speed of concurrent jobs. We shown the APIs as follows:
- *initPtable*: specify each block's initial priority value;
- *De_In_Priority*: decide the priority queue for each job based on block.
- *De_Gl_Priority*: decide the globle priority queue based on priority queue of every jobs.
- *Con_processing*: schedule jobs to process graph data based on globle priority queue

The use of APIs is shown in Fig. 9. The fuctions of *initPtable* is invoked when a new job is coming, and the fuctions of *De_In_Priority* and *De_Gl_Priority* can be invoked after each iteration. So, these three fuctions can be easily added to the traditional graph processing systems. But in fuction of *Con_processing*, it needs to be combined with the computing model of the original system. Because it need to schedule jobs and blocks of graph data.

In order to support prioritized iteration more effectively, we implement the graph algorithm in delta-based accumulative iterative computation, such as PageRank algorithm is updated as follows:

$$\begin{cases} P_j^k = P_j^{k-1} + \Delta P_j^k \\ \Delta P_j^{k+1} = \sum_{\{i|(i \to j) \in E\}} d * \frac{\Delta P_i^k}{|N(i)|} \end{cases} \quad (3)$$

where $P_j^k$ is the value of node j after k iterations, d is damping factor, $\Delta P_j^k$ is the change from $P_j^{k-1}$ to $P_j^k$, $|N(i)|$ is the number of out-edge of node i, and E is the set of in-edge.

## 5 OPTIMIZATION

### 5.1 Optimal Length

The length of priority queue is very important because of which is critical in determining how many computations are required for job convergence. If the priority queue length is too short, the less amount of computation of each iteration will be. Nevertheless, the shorter priority queue will lead to the number of iterations increases greatly and causes more frequent priority queue maintenance operations. On the other hand, if the priority queue length is too large, then the amount of computation per iteration will be increased. Hence, it is not much different from traditional methods. Instead, maintaining priority queues requires overhead.

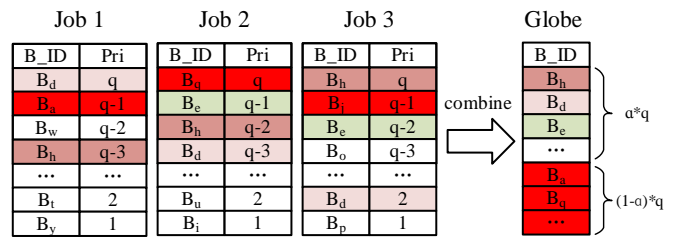

Fig. 7: Generate the globe priority queue.

In order to determine the priority queue length q based on block scheduling, we first comprehend the priority queue length Q based on nodes. In PrIter, the priority queue length for each individual task is selected to $Q = C * \sqrt{V_N}$, $V_N$ is the number of nodes in the graph structure, C is a constant and be 100 by default. The selection formula of Q can obtain better convergence acceleration effect, so we hope to apply this method to the prioritized iteration based on block scheduling. Hence, we set q to be $q = Q/V_B$. If the number of blocks is $B_N$, then $V_N = B_N \times V_B$. Then we have the optimal length of globe priority queue:

$$q = C * B_N/\sqrt{V_N} \quad (4)$$

and also set the C default to 100.

## 6 RELATED WORK

Distributed graph processing systems store the graph data to each node by partition and exploiting the powerful computation resources of clusters. However, due to the strong correlation between graph data and the characteristic of power-law distribution, there will be lots of synchronization overhead and load imbalance in distributed environments. These challenges are the bottleneck of distributed graph processing system.

At the same time, single-machine graph processing systems are able to handle graphs with billions of edges by using secondary storage efficiently, and largely eliminate all the challenges of distributed frameworks. [In this kind of systems, the graph data partitions are processed in memory in order, and the partitions of remaining data are stored in the secondary storage. Therefore, users do not require powerful distributed clusters when using such systems, as well as do not need the ability to managing and tuning distributed clusters. With the development of hardware technology in recent years, limited computing power and system resources, the bottleneck of single-machine, can be effectively alleviated. Firstly, many commodity single-node servers can easily extend memory to hundreds of GBs to TBs. Secondly, current accelerators have much higher massive parallelism and memory access bandwidth than traditional CPUs, which has the potential to offer high-performance graph processing. Such as, Garaph , CuSha , etc., based on GPU, and etc. based on FPGA.

## 7 CONCLUSION

With the rapidly growing demand of graph processing in the real scene, we have to efficiently handle massive concurrent jobs. In order to solve the problems existing in the existing methods, we proposed two-level scheduling strategies in this paper, which enable enhance the efficiency of the data access and accelerate the convergence speed of concurrent jobs. Firstly, correlation aware job scheduling allows concurrent jobs process same graph data in Cache, which enable enhance the Cache hit rate. Secondly, multiple priority data scheduling provides the support of prioritized iteration for concurrent jobs, which according to global priority generated by individual priority of each job.

## REFERENCES

[1] Xue J, Yang Z, Qu Z, et al. Seraph: an efficient, low-cost system for concurrent graph processing[M]. 2014.

[2] Zhang Y, Gao Q, Gao L, et al. PrIter:a distributed framework for prioritized iterative computations[C]// ACM Symposium on Cloud Computing. ACM, 2011:1-14.

[3] X. Liao, H. Jin, Y. Liu, L. Ni, and D. Deng, "Anysee: Peer-to-peer live streaming," in Proc. INFOCOM, 2006:1-10.

[4] P. Wang, K. Zhang, R. Chen, H. Chen, and H. Guan, "Replication-based fault-tolerance for large-scale graph processing," in 2014 44th Annual IEEE/IFIP International Conference on Dependable Systems and Networks. IEEE, 2014, pp. 562–573.

[5] Z.Khayyat,K.Awara,A.Alonazi,H.Jamjoom,D.Williams,and P. Kalnis, "Mizan: a system for dynamic load balancing in large-scale graph processing," in Proceedings of the 8th ACM European

[6] Malewicz G, Austern M H, Bik A J C, et al. Pregel: a system for large-scale graph processing[C]// ACM SIGMOD International Conference on Management of Data. ACM, 2010:135-146.

[7] Y.Zhao,K.Yoshigoe,M.Xie,S.Zhou,R.Seker,andJ.Bian, "Lightgraph: Lighten communication in distributed graph-parallel processing," in 2014 IEEE International Congress on Big Data. IEEE, 2014, pp. 717–724.

[8] Gonzalez J E, Low Y, Gu H, et al. PowerGraph: distributed graph-parallel computation on natural graphs[C]// Usenix Conference on Operating Systems Design and Implementation. USENIX Association, 2012:17-30.

[9] Gonzalez J E, Xin R S, Dave A, et al. GraphX: graph processing in a distributed dataflow framework[C]//Usenix Conference on Operating Systems Design and Implementation. USENIX Association, 2014:599-613.

[10] Yucheng Low, Joseph E. Gonzalez, Aapo Kyrola, et al. GraphLab: A New Framework For Parallel Machine Learning[J]. Computer Science, 2010.

[11] Kyrola A, Blelloch G, Guestrin C. GraphChi: large-scale graph computation on just a PC[C]// Usenix Conference on Operating Systems Design and Implementation. 2012:31-46.